\documentclass[twocolumn,floatfix,prl]{revtex4-2}%
\usepackage[dvipdfmx]{graphicx}%
\usepackage{amsmath}%
\usepackage{amsfonts}%
\usepackage{amssymb}
\usepackage{bm}
\usepackage{color}

\def\e{{\epsilon}}
\def\k{{ {\bm k} }}

\def\q{{ {\bm q} }}

\def\0{{ {\bm 0} }}
\def\R{{ {\bm R} }}

\def\r{{ {\bm r} }}

\allowdisplaybreaks[4]

\begin{document}
\title{\textcolor{black}{
Real-space Loop Current Pattern in Time-reversal-symmetry Breaking Phase in
Kagome Metals}}
\author{
Koki Shimura$^1$, Rina Tazai$^2$, Youichi Yamakawa$^1$, Seiichiro Onari$^1$, and Hiroshi Kontani$^1$
}

\date{\today }

\begin{abstract}
The charge loop current (cLC) state has attracted increasing attention in kagome metals. 
Here, we calculate the currents along the nearest sites $i$ and $j$, $J_{i,j}$, induced by the cLC order that is the imaginary and odd-parity hopping integral modulation $\delta t_{i,j}$.
We reveal that the magnitude of $J_{i,j}$ strongly depends on the nearest sites $i$ and $j$ in the $2\times2$ cLC state, where $\eta\equiv |\delta t_{i,j}|$ is equivalent for all nearest sites. 
The obtained $J_{i,j}$ becomes large near the van-Hove singularity (vHS) filling ($n\sim n_{\rm vHS}$) even when $\eta$ is fixed.
Interestingly, the obtained $J_{i,j}$ exhibits the logarithmic divergence behavior at low temperatures for $n\sim n_{\rm vHS}$ with a fixed $\eta$,
by reflecting the vHS points that are the characteristic of kagome metals.
The present study provides useful information for local electronic state measurements, such as the site-selective NMR and STM experiments.
\end{abstract}

\address{
$^1$Department of Physics, Nagoya University,
Furo-cho, Nagoya 464-8602, Japan\\
$^2$ Yukawa Institute for Theoretical Physics, Kyoto University, Kyoto 606-8502, Japan.
}
 

\sloppy

\maketitle

\textit{Introduction} - 
In recent years, kagome lattice metals have become one of central topics 
in the field of strongly correlated electron systems. 
Multiple exotic metallic phase transitions, 
such as the star-of-David bond-order (BO), superconducting (SC) state,
and time-reversal-symmetry-breaking (TRSB) phase without spin polarization, 
are realized thanks to the strong geometrical frustration that prevents 
simple spin and charge orders. 
In $A$V$_3$Sb$_5$ ($A$=Cs,Rb,K), the $2 \times 2$ BO
appears at $T_{\rm BO}\approx100$K
\cite{STM1,kagome-exp4}.
Inside the BO phase, the SC state appears at $T_{\rm SC} \approx 1-10$K 
\cite{kagome-exp1,kagome-exp2,kagome-P-Tc1,kagome-P-Tc2,kagome-P-Tc3}.
In addition, the electronic nematic state has been observed inside 
\cite{elastoresistance-kagome,birefringence-kagome, STM2} 
and outside \cite{Asaba-torque} of the BO phase.
To explain the observed nematicity,
the single-${\bm Q}$ odd-parity BO \cite{Tazai-Morb}
and $E_g$ symmetry nematic BO \cite{Huang-mix} 
have been proposed theoretically.

To understand these exotic electronic states in kagome metals,
various theoretical studies have been performed, based on
the extended mean-field theory \cite{Neupert2021,Nandkishore},
the functional renormalization group (RG) theory \cite{Thomale2013,SMFRG},
the parquet RG theory \cite{Balents2021,Nat-g-ology},
and the density-wave (DW) equation analysis with the higher-order vertex corrections (VCs) \cite{Tazai-kagome}.
Both the RG theory and the DW equation theory
have been successfully applied to Fe-based \cite{Onari-SCVCS,Onari-form,Yamakawa-PRX,Kontani-rev2,Tazai-LW,Chubukov-RG}
and cuprate superconductors \cite{Tsuchiizu-Cu}.
The nematic order and the BO in these systems 
originate from the higher-order VCs
\cite{Kontani-rev2}.
Recently, the present authors proposed
a unified explanation for the BO and the heavily anisotropic $s$ wave SC state
in kagome metals based on the DW equation analysis
\cite{Tazai-kagome}.
This analysis also predicts the emergence of the electronic toroidal quadrapole order in CsTi$_3$Bi$_5$
\cite{Ti-kagome}.
However, theoretical understanding for the TRSB state has been limited until recently.

Recently, the TRSB state has been observed by various experimental methods.
Inside the BO phase, the TRSB has been observed by
the STM \cite{STM1},
the $\mu$SR \cite{muSR4-Cs,muSR5-Rb,muSR2-K}, 
the anomalous Hall effect \cite{AHE1, AHE2}, 
and the magnetochiral anisotropy (eMChA) \cite{Moll-eMChA,Moll-hz} measurements.
The spontaneous charge loop-current (cLC) order is  
a natural candidate of the TRSB order.
In kagome metals, the $2\times2$ cLC order has finite orbital magnetization $M_{\rm orb}$;
see Ref. \cite{Tazai-Morb}.
References \cite{Moll-eMChA,Moll-hz} report that 
the TRSB domains with random chiralities are detwinned by the magnetic field $h_z\sim1{\rm T}$,
and $T_{\rm TRSB}$ is drastically enhanced by small $h_z$ as well as the uniaxial pressure $\e$.
Such drastic $h_z$- and $\e$-dependences of the TRSB states are naturally explained
based on the Ginzburg-Landau (GL) theory under $(h_z,\e)$ in 
Ref. \cite{Tazai-Morb}, owing to the finite $M_{\rm orb}$.
Very interestingly,  
the TRSB state outside the BO phase ($T_{\rm TRSB}\approx130$K)
has been discovered by the recent magnetic torque measurement \cite{Asaba-torque}.
The single ${\bm Q}$ cLC order ($M_{\rm orb}=0$) is a natural candidate.
Actually, the cLC order can emerge above $T_{\rm BO}$ theoretically \cite{Tazai-kagome2}.

The cLC order was originally studied in cuprate superconductors
\cite{Affleck,Varma-LC,Andersen-LC}.
The cLC order is defined as the imaginary and odd-parity hopping integral modulation between sites $i$ and $j$,
$\delta t_{i,j}$ \cite{Tazai-kagome,Tazai-kagome2,Kontani-spinloop,Tazai-cLC}. 
The Berry curvature due to $\delta t_{i,j}$ 
gives rise to the spontaneous current \cite{Haldane}. 
To understand the microscopic origin of the cLC order in kagome metals,
various theoretical studies have been performed, such as the 
mean-field theory \cite{Neupert2021,Nandkishore},
the parquet RG theory \cite{Balents2021,Nat-g-ology},
and the DW equation analysis \cite{Tazai-kagome2}.
Importantly, the BO fluctuations in kagome metals mediate not only the SC pairing,
but also the cLC order that is the TRSB particle-hole pair condensation;
see Ref. \cite{Tazai-kagome2}. 
Therefore, the cLC state is naturally expected to occur in kagome metals with the BO instability.
Recently, the competition between the BO, cLC and the SC states is intensively studied 
based on the GL theories \cite{Balents2021,Tazai-Morb,Thomale-GL,Fernandes-GL}.

In the cLC state in kagome metal model, 
the uniform orbital magnetization $M_{\rm orb}$
was recently studied in Ref. \cite{Tazai-Morb}.
However, the nanoscale charge current flowing between the nearest bonds,
$J_{i,j}$, has not been performed to our knowledge.
The knowledge of $J_{i,j}$ is 
significant to understand the local magnetization measurements,
such as the $\mu$SR \cite{muSR4-Cs,muSR5-Rb,muSR2-K}
and the NMR \cite{Zheng} measurements.

In this Letter, 
we analyze the \textcolor{black}{ current $J_{i,j}$ in real space} in the $2\times2$ cLC order phase.
We reveal that the magnitude of $J_{i,j}$ strongly depends on the nearest sites $(i,j)$.
The obtained $J_{i,j}$ becomes large near the van-Hove singularity (vHS) filling ($n\sim n_{\rm vHS}$).
Interestingly, $J_{i,j}$ exhibits the logarithmic divergence behavior at low temperatures for $n\sim n_{\rm vHS}$,
by reflecting the vHS points that are the characteristic of kagome metals.
The present study provides useful information for local electronic state measurements, such as the site-selective NMR and STM experiments.

\begin{figure}[!htb]
\includegraphics[width=.99\linewidth]{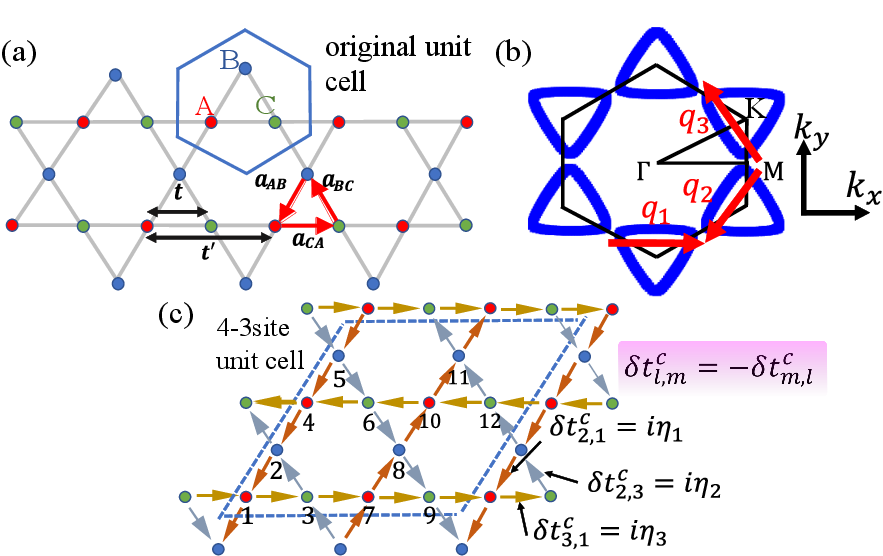}
\caption{(Color online) 
(a) Tight-binding model of the kagome lattice model. (b) FS of the 3-site kagome model at $n=5.0$. (c) The $3\bm{Q}$ cLC $\bm{\eta} = (\eta,\eta,\eta)/\sqrt{3}$ depicted in the kagome lattice.}
\label{fig:kagome-model}
\end{figure}

\textit{Model Hamiltonian with $3\bm{Q}$ cLC Order} - First, we
\textcolor{black}{present} a 3-site kagome lattice model
\textcolor{black}{as shown} in Fig. \ref{fig:kagome-model}
(a). \textcolor{black}{The} nearest-neighbor hopping integral is set to $t
= -0.5$[eV]\cite{Thomale2021}. To \textcolor{black}{avoid} perfect nesting, a
\textcolor{black}{3rd-neighbor hopping integral along sides of two triangles is introduced with $t' = -0.02$[eV] as shown in
Fig. \ref{fig:kagome-model} (a)}. \textcolor{black}{The unit cell contains three sites, labeled A, B, and C.} The vectors between nearest-neighboring sites are defined as $\bm{a}_{AB},\bm{a}_{BC}$, and $\bm{a}_{CA}$. Fig. \ref{fig:kagome-model} (b) shows the Fermi surface (FS) of the 3-site kagome lattice model and \textcolor{black}{the} nesting vectors $\q_n$ $(n=1,2,3)$.Three $\q_n$ vectors connect vHS at the M point.
 
\textcolor{black}{When the BO or cLC with the nonzero wavevectors $\q_n$
appears in the kagome lattice}, the translational symmetry of the system
is violated. \textcolor{black}{In the case of a $3\bm{Q}$ order, the unit
cell is extended to include 12 sites.}
Fig. 1 (c) shows the $3\bm{Q}$ \textcolor{black}{cLC} with the order
parameter $\bm{\eta} = (\eta_1,\eta_2,\eta_3)$,
\textcolor{black}{where $\eta_n$ is current order parameter
with $\q_n$.}
We assign $\delta t_{i,j}^{c}=i\eta_1$ \ $[\delta t_{i,j}^{c}=-i\eta_1]$ for $(l,m)=(1,2),(2,4),(4,5),(5,1)$ $[(l,m)=(7,8),(8,10),(10,11),(11,7)]$, $\delta t_{i,j}^{c} = i\eta_2$ \ $[\delta t_{i,j}^{c}=-i\eta_2 ]$ for $(l,m)=(2,3),(3,11),(11,12),(12,2)$$[(l,m)=(5,6),(6,8),(8,9),(9,5)]$, and $\delta t_{i,j}^{c}=i\eta_3 \ [\delta t_{i,j}^{c}=-i\eta_3 ]$ for $(l,m)=(4,6),(6,10),(10,12),(12,4)$$[(l,m)=(1,3),(3,7),(7,9),(9,1)]$. Here,  $\delta t^c_{i,j}$ is the purely imaginary ($(\delta t_{i,j}^c)^{*} = -\delta t_{i,j}^c$) symmetry breaking term of the hopping integral, which is odd parity ($\delta t^c_{i,j} = -\delta t^c_{j,i}$) \cite{Tazai-kagome2}.

The Hamiltonian with the current order is given by
\begin{equation}
	\hat{H} = \sum_{\k,l,m}h_{lm}(\k)c^{\dagger}_{\k,l} c_{\k,m},
\end{equation}
\textcolor{black}{$l$ and $m$ indicate the site numbers in the unit cell, with $l,m = 1 \sim 12$.
$h_{lm}(\k)$ represents the Fourier transformation of $t_{i,j} =
t^{(0)}_{i,j} + \delta t^c_{i,j}$, where $t^{(0)}_{i,j}$ is the original hopping integral.}
\textcolor{black}{An index $i=\{\alpha, l\}$ is employed, where $\alpha$ denotes the unit cell.} We introduce the notation $\R_i=\R_{\alpha}+\r_l$ to designate the position of atoms, where $\R_{\alpha}$ is the position of the unit cell and $\r_l$ is the relative position of the atom within the unit cell. In the 12-site kagome lattice model, the FS folds, and the M and $\Gamma$ points in the lattice become equivalent. Therefore, the wave vector of the cLC order is $\q_n = \bm{0}$ for $n = 1, 2, 3$ within folded BZ.

Here, we introduce the form factor of the cLC, which is given by the Fourier transformation of $\delta t^c_{i,j}$ shown in Fig. \ref{fig:kagome-model} (c). In the 3-site kagome lattice model, the cLC form factor between the nearest site $S$ and site $S'$ ($S,S'=A,B,C$) is given as $f_{SS'}^{\q_n}(\k)= 2\sin(\k \cdot \bm{a}_{SS'})$, where inter-site vector $\bm{a}_{SS'}$ and the wavevector $\q_n$ are shown in Fig. \ref{fig:kagome-model} (a) and (b), respectively.
Here, $n=1,2,3$ for $\left\{S,S'\right\}=\left\{A,B\right\}, \left\{B,C\right\}, \left\{C,A\right\}$, respectively.
The set of the current order functions is given by $(\eta_1f_{AB},\eta_2f_{BC},\eta_3f_{CA})$, where $\eta_n$ is the $n$th cLC order parameter.
A more detailed explanation is presented in Ref. [20].
Hereafter, we set $\eta_1 = \eta_2 = \eta_3 = \eta$ in the numerical study.
We can also present the cLC form factor in the 12-site model, ${\tilde f}_{lm}(k)$, where $l,m=1 \sim 12$ \cite{Tazai-Morb}. For example, ${\tilde f}_{12}(\k) = ie^{-i\bm{k}\cdot\bm{a}_{12}}$. In this representation, the cLC order wavevector becomes uniform $(\q=\bm{0})$, so the incoming and outgoing momenta of ${\tilde f}_{lm}(\k)$ are the same.
Here, we introduce the $12 \times 12$ matrix expression of the current order function $[{\hat C}_{\k}]_{l,m}\equiv {\tilde f}_{lm}(\k)$, which is convenient in the present numerical study. Here, $[{\hat C}_{\k}]_{l,m}\ne0$ only when $(l, m)$ is the pair of the nearest sublattice.

The $3\bm{Q}$ cLC form factor of kagome metals is derived microscopically from the analysis of the DW equation with the BO fluctuation exchange term as the kernel, as shown in ref. \cite{Tazai-kagome2}. The $\delta t_{i,j}$ derived in ref. \cite{Tazai-kagome2} has the same absolute value for all nearest-neighbor bonds $(i,j)$, meaning $|\delta t_{i,j}^c| = \eta$. 
It is worth noting that the $\delta t_{i,j}$ derived in ref. \cite{Tazai-kagome2} also has long-range components, but for simplicity in this paper, we consider only the nearest-neighbor components.

\begin{figure}[!htb]
\includegraphics[width=.99\linewidth]{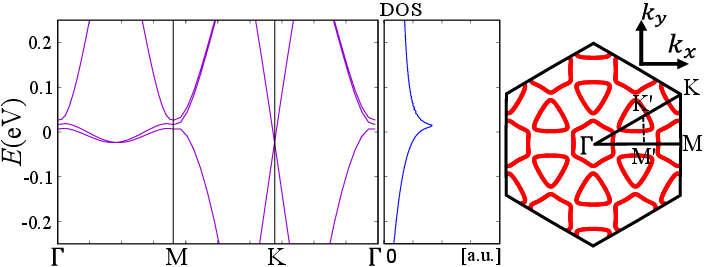}
\caption{(Color online) 
Band structure, DOS, and FS of the 12-site kagome lattice model with $\eta = 0.1$.
}
\label{fig:band-fermi}
\end{figure}

We show the band structure, density of states (DOS), and FS of the 12-site kagome lattice model that cLC order is introduced at $\eta = 0.1$ in Fig \ref{fig:band-fermi}. In the 12-site model, vHS exists at the M-point, leading to a large DOS. The introduction of cLC order resolves the strong degeneracy near the M point, while the large magnitude of the DOS remains.

Next, to calculate the charge current in the cLC phase, we introduce the current order operator from Heisenberg eq.
$\dot{n}_i = -i[n_i,H] = -i\left(\sum_{j}t_{i,j}c_{i}^{\dagger} c_{j} - \sum_{j}t_{j,i}c_{j}^{\dagger} c_i\right)$. 
Then, the current operator is derived from the continuity equation as
\begin{eqnarray}
	j_{i,j} = -i(t_{i,j}c^{\dagger}_i c_j - (i \leftrightarrow j)).
	\label{eqn:current operator}
\end{eqnarray}
By taking the expectation value with respect to grand canonical ensemble, the current from site $j$ to site $i$ is given as
\begin{eqnarray}
	J_{i,j} = \langle j_{i,j} \rangle = it_{i,j}g_{j,i}- (i \leftrightarrow j),
	\label{eqn:current form}    
\end{eqnarray}
where the equal time Green function is given as $g_{i,j} = \lim_{u \rightarrow -0}\langle T c^{\dagger}_i(u)c_j(0) \rangle$. $\langle \cdots \rangle$ represents the mean value for the grand canonical ensemble. 
The Green function in momentum representation is
\begin{eqnarray}
G_{l,m}(\k,i\epsilon_n) = ((i\epsilon_n + \mu)\hat{\bm{1}} - \hat{h}(\k))^{-1}_{l,m}.
\label{eqn:Greenfunction_momentum}    
\end{eqnarray}
We derive the equal time Green function by the Fourier transform of Eq. (\ref{eqn:Greenfunction_momentum}) as $g_{i,j} = T\sum_{n,\k}G_{l,m}(\k,i\epsilon_n)e^{i\k \cdot (\R_\alpha-\R_\beta)}e^{i\epsilon_n\delta}$. Note that the factor $e^{i\epsilon_n\delta}$ is not necessary for $l \neq m$. $\epsilon_n = (2n-1)\pi T$ is the fermion Matsubara frequency. Here n is integer.


\begin{figure}[!htb]
\includegraphics[width=.99\linewidth]{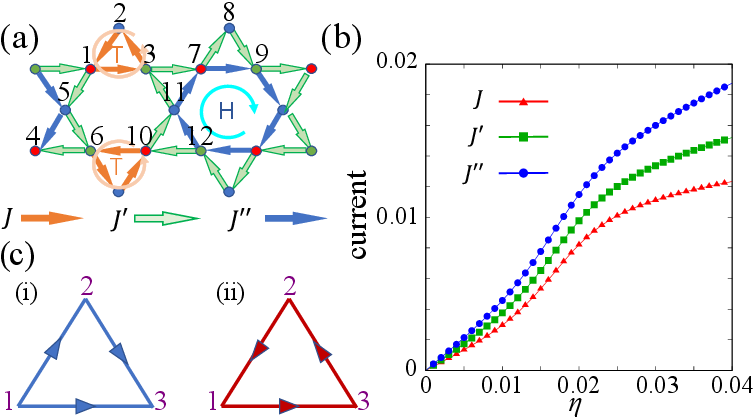}
\caption{(Color online) (a) Obtained distribution of the charge currents. Here, $J (J'') $ is the current along the triangular (hexagonal) loop where the direction of each cLC order is aligned. The current on other bonds is given as $J'$.(b) Obtained currents. (c) 3-site cluster models with different cLC orders.  The circulating current in $\rm(\hspace{.18em}i\hspace{.18em})$ differs from that in $\rm(\hspace{.08em}ii\hspace{.08em})$; see the text.}
\label{fig:3-site-cluster}
\end{figure}

\textit{Analysis of Current} -
Now, we present the results of the current distribution induced by the cLC orders. Fig. \ref{fig:3-site-cluster} (a) \textcolor{black}{shows} the real-space current distribution within the kagome lattice. The obtained currents as a function of $\eta$ are shown in Fig. \ref{fig:3-site-cluster} (b) ($T = 2.5$meV, $n = 5.0$) . As $\eta$ increases, the difference in the magnitude of the currents is enlarged. 

To understand the obtained nontrivial relation between $J_{i,j}$ and $\delta t_{i,j}$, here we analyze a simple 3-site cluster model with cLC order oriented in different directions. Fig. \ref{fig:3-site-cluster} (c) represents two 3-site cluster models, $\rm(\hspace{.18em}i\hspace{.18em})$ and $\rm(\hspace{.08em}ii\hspace{.08em})$. In the case of $\rm(\hspace{.18em}i\hspace{.18em})$, $\delta t^c_{ij}$ is given as $\delta t^c_{21} = \delta t^c_{31} = \delta t^c_{32} = i\eta$. The current $J^1_{1,2}$ that flows in model $\rm(\hspace{.18em}i\hspace{.18em})$ is derived as $J^1_{1,2} = -2\eta(t^2+\eta^2)A$, where $A = T\sum_n \left[(i\epsilon_n + \mu)^3 + 3(t^2+\eta^2)(i\epsilon_n + \mu)
+2t(t^2+\eta^2)\right]^{-1}$.
In the case of $\rm(\hspace{.18em}ii\hspace{.18em})$,  $\delta t^c_{ij}$ is given as $\delta t^c_{12} = \delta t^c_{23} = \delta t^c_{31} = i\eta $. Then, the current $J^2_{1,2}$ is calculated as $J^2_{1,2} = 6\eta(t^2-\frac{\eta^2}{3})B$, where $B = T\sum_n \left[(i\epsilon_n + \mu)^3 + 3(t^2+\eta^2)(i\epsilon_n + \mu) 
\right.$ $\left.+2t(t^2-3\eta^2)\right]^{-1}$.
Thus, different magnitudes of current emerge even when the same order parameter $\eta$ is provided in three directions $\q_1, \q_2, \q_3$.  It holds that $|J_{12}^m| = |J_{23}^m| = |J_{31}^m|$ for each $m = 1,2$.
In the case that $|a| \ll |t|$, $|J_{12}^2| = 3|J_{12}^1|$ holds. In this finite site model, the current converges to zero at $T=0$ and $T \rightarrow \infty$.

The mathematical reason for the difference in current between $\rm(\hspace{.18em}i\hspace{.18em})$ and $\rm(\hspace{.18em}ii\hspace{.18em})$ is that \textcolor{black}{the $J_{1,2}$ given in Eq. (\ref{eqn:current form}) includes contributions not only from $\delta t_{1,2}^c$ but also from $\delta t_{2,3}^c$ and $\delta t_{3,1}^c$,} which are included in $g_{1,2}$. When the phases of $\delta t_{i,j}^c$ on the closed loop are aligned,  i.e., $\delta t_{1,2}^c = \delta t_{2,3}^c = \delta t_{3,1}^c$, the induced current $|J_{i,j}|$ becomes the largest. This consideration leads us to understand why $J$, $J'$, and $J''$ are different in kagome metals as shown in Fig. \ref{fig:3-site-cluster} (a)-(b).

\begin{figure}[!htb]
\includegraphics[width=0.99\linewidth]{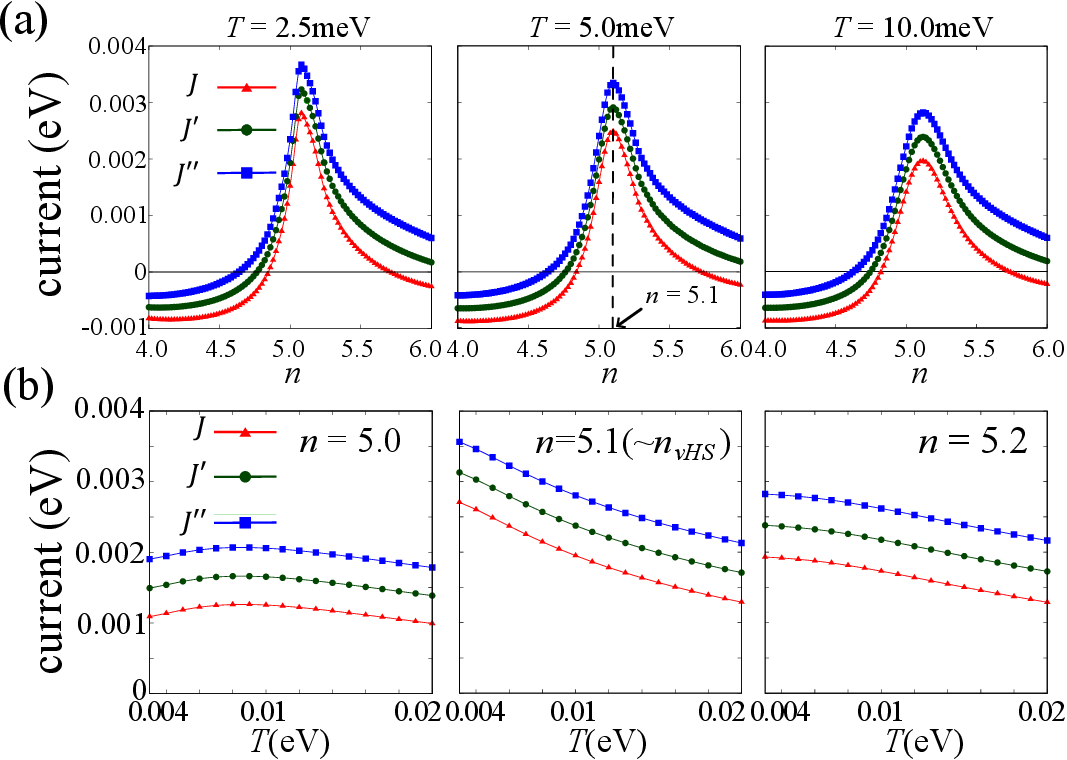}
\caption{(Color online) (a) n-dependence of the currents at $T = 2.5-10.0$meV. (b) T-dependence of the currents at $n = 5.0-5.2$.}
\label{fig:third-current}
\end{figure}


Next, we analyze the filling and temperature dependence of the current. The filling dependence of the current is depicted in Fig. \ref{fig:third-current} (a). It shows that the current reaches a maximum near the vHS filling. Fig. \ref{fig:third-current} (b) illustrates the temperature dependence of the current. The magnitude of the current increases at the low-temperatures, with a pronounced increase near the vHS filling.

\begin{figure}[!htb]
\includegraphics[width=0.99\linewidth]{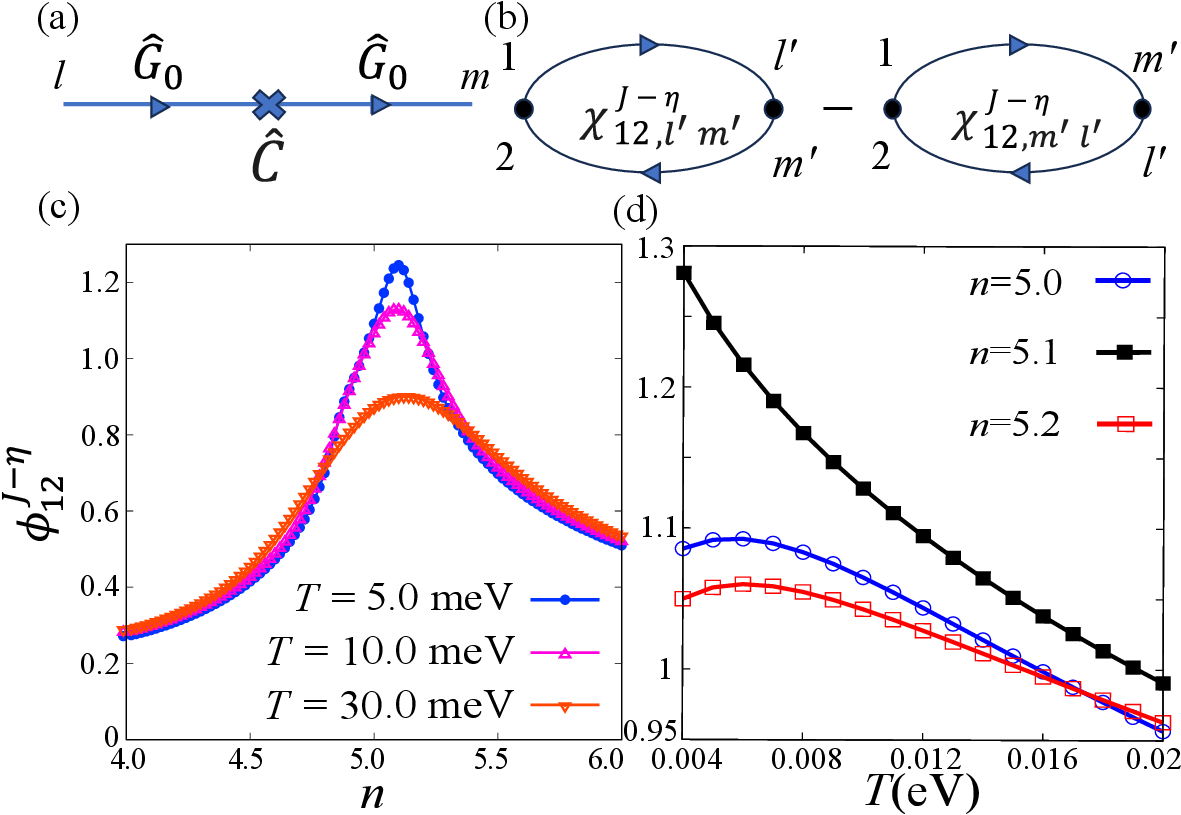}
\caption{(Color online) (a) Diagrammatic representation of the first-order term $\left(\hat{G}^{(0)}\eta \hat{C}\hat{G}^{(0)}\right)_{m,l}$. (b) Diagrammatic representation of the sum of susceptibility $\chi_{12,l'm'}^{J-\eta} - \chi_{12,m'l'}^{J-\eta}$. (c)$n$-dependence of $\phi_{12}^{J-\eta}$ at $T = 5.0, 10, 30$ meV. (d) Temperature dependence of $\phi_{12}^{J-\eta}$ at $n=5.0-5.2$}.
\label{fig:chi-sum-v2}
\end{figure}

\textit{Susceptibility of the Current} - 
Next stage, we consider susceptibility that describes the behavior of the current. 
First, we expand the Green function in terms of the order parameter $\eta$ as $\hat{G}\left(\k,i\epsilon_n\right)\sim \hat{G}^{(0)}\left(\k,i\epsilon_n\right) + \hat{G}^{(0)}\left(\k,i\epsilon_n\right)\eta {\hat{C}}_{\k}\hat{G}^{(0)}\left(\k,i\epsilon_n\right)$, where $\hat{G}^{(0)}(\k,i\epsilon_n)$ is the Green function with $\eta = 0$. The contribution from the term $\hat{G}^{(0)}(\k,i\epsilon_n)\eta {\hat{C}}_{\k}\hat{G}^{(0)}(\k,i\epsilon_n)$ to the current is dominant. The diagrammatic representations of this term is shown in Fig. \ref{fig:chi-sum-v2} (a). 
Meanwhile, the static irreducible susceptibility is
\begin{eqnarray}
    &&\chi^{J-\eta}_{lml'm'}(\q)\notag\\
	&&= -\frac{T}{N}\sum_{\k,n}G^{(0)}_{ll'}(\k+\q,i\epsilon_n)G^{(0)}_{m'm}(\k,i\epsilon_n).
	\label{eqn:representation-of-chi}
\end{eqnarray}
As an example of form factor, $[\hat{C}_{{\k}}]_{12} = ie^{i\k \cdot \bm{a}_{AB}}$, and $[\hat{C}_{\k}]_{21} = -ie^{i\k\cdot \bm{a}_{BA}}$. Thus, at $\k=0$ which corresponds to $\Gamma$ point, $[\hat{C}_{\bm{0}}]_{12} = i$ holds. Here, we recall that sites $1$,$4$,$7$, and $10$ belong to the A site, while sites $2$,$5$,$8$, and $11$ belong to B site. In the case of the 12-site kagome lattice model, the form factor exhibits the largest value at the $\Gamma$ point. Therefore, after approximating it as $[\hat{C}_{\k}]_{lm} \sim [\hat{C}_{\bm{0}}]_{lm}$, \textcolor{black}{$J_{1,2}$ is given as
\begin{eqnarray}
	J_{1,2} &&\propto T\sum_{n,l',m',\k} [G^{(0)}]^{1,l}\eta [\hat{C}_{\k}]_{l'm'}[G^{(0)}]^{m',2}e^{i\k \cdot (\R_\alpha-\R_\beta)} \notag \\ && \sim \sum_{l',m' = 1 - 12} \chi_{12,l'm'}^{J-\eta}(\bm{0})[\hat{C}_{\bm{0}}]_{l'm'}\eta.
	\label{eqn:current-chi}
\end{eqnarray}
}
Importantly, $\chi_{12,l'm'}^{J-\eta}(\bm{0})$ becomes large when
$\left\{ l',m' \right\} = \left\{1,2\right\}, \left\{4,5\right\},
\left\{7,8\right\},\left\{10,11\right\}$, by reflecting the sublattice
interference of the pure-type FS \cite{Tazai-kagome}. \textcolor{black}{When we focus on
the contribution from $\left\{ l',m' \right\} =
\left\{1,2\right\},\left\{2,1\right\}$, $J_{1,2}$ in
Eq. (\ref{eqn:current-chi}) is proportional to $\chi_{12,12}^{J-\eta} -
\chi_{12,21}^{J-\eta}$ due to the odd-parity form factor.}

\textcolor{black}{Figure \ref{fig:chi-sum-v2} (b) shows diagrammatic representations of $\chi_{12,l'm'}^{J-\eta} - \chi_{12,m'l'}^{J-\eta}$.} Taking account of the form factors depicted in Fig. \ref{fig:kagome-model} (c), the summation of contributions from $\left\{l,m\right\} = \left\{1,2\right\}$, $\left\{4,5\right\}$, $\left\{7,8\right\}$, and $\left\{10,11\right\}$ leads to $\phi_{12}^{J-\eta} = [\chi_{12,12}^{J-\eta}-\chi_{12,21}^{J-\eta} + \chi_{12,42}^{J-\eta}-\chi_{12,24}^{J-\eta} + \chi_{12,45}^{J-\eta}-\chi_{12,54}^{J-\eta} + \chi_{12,14}^{J-\eta}-\chi_{12,41}^{J-\eta}] - [\chi_{12,78}^{J-\eta}-\chi_{12,87}^{J-\eta} + \chi_{12,10\ 8}^{J-\eta}-\chi_{12,8\ 10}^{J-\eta} + \chi_{12,10\ 11}^{J-\eta}-\chi_{12,11\ 10}^{J-\eta} + \chi_{12,7\ 10}^{J-\eta}-\chi_{12,10\ 7}^{J-\eta}]
$.

The filling and temperature dependences of $\phi_{12}^{J-\eta}$ are
shown respectively in Fig. \ref{fig:chi-sum-v2} (c) and
(d). $\phi_{12}^{J-\eta}$ reaches its maximum value at the vHS filling
$n_{\rm vHS}\sim 5.1$, with a notable increase in the low temperature
regime. Analytically, at vHS filling, $\phi_{12}^{J-\eta} \propto 1/ E_0
\left( \ln E_0/T\right)^2 + \chi_{0c}$ when $E_0 \sim 0.1$[eV] is the
energy range where the vHS bandstructure is well described by the
quadratic expression, as discussed in Ref. \cite{Balents2021}. Here,
$\chi_{0c}$ is the constraint susceptibility, which excludes the
contribution for $|E| < E_0$ from Eq. (\ref{eqn:representation-of-chi}),
which is defined in SM in Ref. \cite{Tsuchiizu-Orbital}. This
temperature dependence is attributed to the vHS in the kagome lattice
model, in stark contrast to usual metals, where $\phi_{12}^{J-\eta}$
remains constant.

\textcolor{black}{The susceptibility $\phi_{12}^{J-\eta}$ can be obtained experimentally by observing $\eta$ and $J_{ij}$, which in principle may be derived from the local DOS and the local magnetic-field measurements, respectively. This is an interesting future challenge.}

\begin{figure}[!htb]
\includegraphics[width=0.99\linewidth]{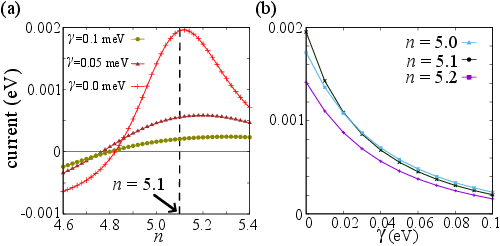}
\caption{(Color online) 
(a) Filling dependence of the current $J^{\rm min}=\min{J,J',J''}$ for various $\gamma$. (b) $\gamma$ dependence of $J^{\rm min}$ for $n = 5.0-5.2$.
}
\label{fig:self-energy}
\end{figure}

\textit{Effect of Self-energy} - 
In addition, we analyze how the current is modified by the self-energy. Fig. \ref{fig:self-energy} (a) shows the filling dependence of the current in the presence of self-energy. Fig. \ref{fig:self-energy} (b) shows the obtained damping rate $\gamma$ dependence of current. As $\gamma$ increases, the current is gradually suppressed.

{\textit{Summary and Discussions} 
- In this letter, we employed the Green function method to calculate the real-space current distribution in the $3{\bm Q}$ cLC phase of kagome metal. 
The obtained magnitude of $J_{i,j}$ strongly depends on the nearest sites ($i,j$) even when $|\delta t_{i,j}|$ is constant for any bonds.
The obtained $J_{i,j}$ becomes large near the vHS filling,
while it is suppressed when considering the self-energy. 
Interestingly, $J_{i,j}$ exhibits the logarithmic divergence behavior at low temperatures for $n\sim n_{\rm vHS}$.
The present study provides useful information for local electronic state measurements, such as the site-selective NMR and STM experiments.}

We have employed the Green function method to calculate $J_{i,j}$.
Based on this method, one can calculate the current by including the beyond-mean-field electron correlations, such as the self-energy and the VCs.
This is a great advantage of the present method.





\acknowledgements

This study has been supported by Grants-in-Aid for Scientific
Research from MEXT of Japan (JP20K03858, JP20K22328, JP22K14003, JP23K03299),
and by the Quantum Liquid Crystal
No. JP19H05825 KAKENHI on Innovative Areas from JSPS of Japan.


\end{document}